# Spreadsheet Structure Discovery with Logic Programming

*Jocelyn Paine,*

*http://www.j-paine.org/*

**ABSTRACT**

*Our term "structure discovery" denotes the recovery of structure, such as the grouping of cells, that was intended by a spreadsheet's author but is not explicit in the spreadsheet. We are implementing structure-discovery tools in the logic-programming language Prolog for our spreadsheet analysis program Model Master, by writing grammars for spreadsheet structures. The objective is an "intelligent structure monitor" to run beside Excel, allowing users to reconfigure spreadsheets to the representational needs of the task at hand. This could revolutionise spreadsheet "best practice".*

*We also describe a formulation of spreadsheet reverse-engineering based on "arrows".*

## 1 INTRODUCTION

In our first paper, presented at EuSpRIG 2001 [Paine, 2001], we introduced the Model Master (MM) spreadsheet-description language, and showed how we could build spreadsheets by compiling them from MM programs, with advantages for readability, modularity, and code-reuse. In an MM version of Thomas Grossman's queuing simulation, where each column represented a server, we could change the number of servers simply by altering one constant.

We also introduced decompilation - translating spreadsheets to MM programs – and its benefits for reverse-engineering and error-detection. An MM program is a summary of a spreadsheet's calculations, useful in many ways. For example, Graham Macdonald has suggested that the commercial world would find it valuable in determining whether a spreadsheet accurately reflects a legal contract.

Our second paper, available on the Web [Paine, 2004], took decompilation further, making it rigorous via "spreadsheet algebra" (the phrase should be understood in the same way as, for example, "vector algebra"), which treats spreadsheets as mathematical entities. We implemented functions for operating on these, and an interface resembling those found in computer-algebra systems, which enabled the results to be displayed as MM programs or recompiled into spreadsheets.

One example decompiled one of Ray Panko's spreadsheets and put it through a series of transformations aimed at making it more intelligible. The freshly decompiled version listed cells individually, giving them their original, uninformative names - D1, D3, and so on. As we proceeded with our transformations, we were able to batch related cells together into arrays and give them names derived from neighbouring labels. Our final listing was easier to read and revealed several intentional errors in the spreadsheet.

Our motivation was that there is no "best" form for such displays. To illustrate: we recently posted on the EuSpRIG discussion list a reference to John Raffensperger's spreadsheet style guidelines [Raffensperger]. Louise Pryor replied that these are indeed good from the point of view of a user who is reading the results and changing some inputs. However, they ignore the needs of other types of users, especially those maintaining and updating the spreadsheet. To us, it is obvious that a spreadsheet must be available in different forms, depending on its user's needs. Blueprints, architectural drawings, circuit diagrams and maps respect this: even library catalogues are searchable both by author and by title. Yet spreadsheets lag behind, forcing users to adapt to one and only one way of ordering their information. Spreadsheet algebra banishes such rigidity.

At the end of that paper, we introduced the term "structure discovery" - discovering logically related cells and redescribing the spreadsheet so that they can be seen as a single data structure. In this paper, we explain structure discovery and describe an implementation which allows the user to define and apply heuristics for finding structure in spreadsheets. Our goal is an "intelligent structure monitor" for Excel, allowing spreadsheets to be reconfigured to the representational needs of the task at hand.

The structure-discovery heuristics are logical specifications of the spreadsheet, and should be coded as close to logic as possible. We are doing so using the logic-programming language Prolog. In the rest of this paper, we introduce structure discovery, explain how common structures can be described by patterns or grammars, and discuss our Prolog implementation. We also formulate reverse-engineering of spreadsheets in terms of "arrows".

We believe Prolog has many advantages for spreadsheet research, and so in our original draft, we included a tutorial on Prolog. On the advice of a referee, this is omitted from the final version appearing in the EuSpRIG 2004 proceedings. However, it can be found in the fuller Web version of our paper, at http://www.j-paine.org/spreadsheet_structure_discovery.html.

It should be noted that although this paper is concerned with one particular piece of software, namely MM, it will still interest other researchers. Structure discovery is necessary whenever one wants to try and describe the spreadsheet so as to tell more about the author's intentions - and hence have more chance of detecting errors - than the spreadsheet itself does. Much of our Prolog analysis code is independent of MM, as is the notion of translating logical specifications directly into Prolog. And our reverse-engineering formalism applies to any attempt to redescribe a spreadsheet as a collection of tables or arrays that, while mapping onto many (perhaps non-neighbouring) cells on the spreadsheet, are nevertheless to be regarded as single data structures.

## 2 STRUCTURE DISCOVERY

We begin with an analogy from the world of conventional programming languages. When a compiler compiles an IF statement, it does so using labels and jumps. The statement:

```
IF Condition THEN
  ThenPart
ELSE
  ElsePart
END IF
```

gets converted into code similar to that below, its exact form depending on the instruction set of the machine:

```
      Condition
      If false, GOTO L1
      ThenPart
      GOTO L2
  L1: ElsePart
  L2:
```

Suppose now that we were to lose the program source. To reconstruct it, we would need a "decompiler", namely a program that reconstructs, from machine code, the source that generated it. (As [Decompilation] describes, many exist.) This IF statement was explicit in the source file, but implicit in the machine code. Spreadsheets too have high-level implicit structure. The difference is that they contain no explicit listing of this structure - it resides only in the author's brain, a location not readily open to public inspection.

Consider as example a three-column Income/Outgoings/Profit spreadsheet, where each cell in the third column computes the difference of its row-mates in the first two:

```
Income Outgoings Profit
              =A2-B2
              =A3-B3
              =A4-B4
```

Logically, the columns are single entities, in the same sense in which the block of machine instructions above is a single high-level statement. Anybody trying to read or maintain the spreadsheet would be greatly helped by knowing both this and what all the third-column calculations have in common. Here, it is obvious; it may not be in larger spreadsheets.

Such analysis is even more vital when spreadsheets complicate their layout for visual effect. Consider a buy-to-rent spreadsheet which performs the same calculation for a number of properties, arraying them in blocks across the sheet, where each block calculates the profit P from rental income R, monthly mortgage payment M, and other costs O.

```
Property 1    Property 2    Property 3
M  O          M  O          M  O
R             R             R
    P             P             P
```

Large spreadsheets of this kind can be confusing, especially when the blocks are split across worksheets.

### 2.1 Representing structure in MM.

How can MM make structure explicit? MM programs contain "attributes" - arrays of elements - related by equations to other attributes. In a freshly decompiled spreadsheet, each attribute is a single cell. Structure discovery entails working out which cells can be grouped together into arrays. Thus, our Income/Outgoings/Profit spreadsheet would look like this at first:

```
<A1 A2 A3 B1 B2 B3 C1 C2 C3>
where A1="Income" B1="Outgoings" C1="Profit"
      C2=A2-B2 C3=A3-B3 C4=A4-B4
```

but could have its attributes grouped and renamed to become more informative:

```
<Income[1..3] Outgoings[1..3] Profit[1..3]>
where Profit[all t] = Income[t] - Outgoings[t]
```

Similarly, the Property one could become

```
<Mortgage{Property1,Property2,Property3}
 OtherCosts{Property1,Property2,Property3}
 Rent{Property1,Property2,Property3}
 Profit{Property1,Property2,Property3} >
where Profit[all p] = Rent[p] - (OtherCosts[p] + Mortgage[p])
```

Of course, the cells do not actually have names. However, one can try guessing plausible names from neighbouring labels. Our spreadsheet algebra interface has functions for this, renaming attributes and redisplaying the resulting program. The user can call these as many times as needed, without committing to any set of names.

### 2.2 How can we discover implicit structure?

We are building up a library of structure-recognition heuristics (or "patterns" or "grammars") that describe common structures and can be automatically matched against spreadsheets. We also allow users to define new patterns, perhaps ones so specific that they apply only to one spreadsheet. This work is described in the following two sections.

Before proceeding, we ought to say that we do not believe this to be the only approach to structure discovery. More powerful methods exist, the ultimate surely being to learn patterns via inductive logic programming [ILP]. We made some other suggestions near the end of our spreadsheet algebra paper [Paine, 2004].

Markus Clermont has written an impressive program for discovering semantically-related regions within spreadsheets [Clermont, 2003]. It searches for evidence of relatedness from a variety of hints; for example, it might propose that cells mentioned together in a cell range or the argument to a SUM should be grouped. Our approach does not pretend to be as comprehensive, but probably shades into Clermont's. Our heuristics are coded in Prolog, which can perform any computation that any other language can, so our patterns could actually invoke any kind of search, including those performed by his program.

## 3 PATTERNS, PATTERN LANGUAGES, AND GRAMMARS

In this section, we explain what we mean by spreadsheet grammars. To develop intuitions, we start with examples from other domains

### 3.1 Pattern examples

**Filename patterns**

Unix and DOS, naturally enough, both allow filenames to be written in full, e.g. `del scratch.tmp`, `copy ss1.xls ss2.xls`. But as a shorthand, they also allow patterns which match sets of filenames. Thus `del *.tmp` deletes all files with the .tmp extension, the asterisk standing for any sequence of characters. So we have symbols (the letters) that stand for themselves, but also symbols that stand for sets of strings.

**Regular expressions**

Regular expressions were invented by mathematician Stephen Cole Kleene, as a notation to manipulate "regular sets", formal descriptions of the behaviour of finite state machines. Today, they form an indispensable part of Unix commands such as "grep", which searches for a string in a set of files. The following examples demonstrate some features:

| | |
|---|---|
| `a` | The letter a |
| `a|b` | An a or b |
| `[a-z]` | Any letter between a and z |
| `([a-z])*` | Any number of such letters |
| `[a-z] ([0-9])*` | One letter followed by any number of digits between 0 and 9 |

Once again, we have symbols that stand for sets of strings. We also have operators that combine patterns into bigger patterns: "|" makes the "or" of two patterns, and "*" following a pattern repeats it indefinitely.

**Snobol patterns**

Snobol [Griswold et. al., 1971] is a string-manipulation language written by David Farber, Ralph Griswold, and I. Polonsky of Bell Labs - and, incidentally, with a silly derivation for its name: StriNg Oriented and symBOlic Language. As these examples, which do the same as the regular expressions, demonstrate, there is more than one way to write patterns:

```
"a"
"a" | "b"
any("abcdefghijklmnopqrstuvwxyz")
arbno( any("abcdefghijklmnopqrstuvwxyz") )
any("abcdefghijklmnopqrstuvwxyz") arbno( any("0123456789") )
```

### 3.2 Grammars

All the above patterns are grammars. A grammar is a formal definition of the syntactic structure of a language, normally written as rules which specify the structure of "phrases" in the language. Each rule has a left-hand side naming a syntactic category (for example, "noun phrase" or "verb" below) and a right-hand side defining it. The right-hand side can contain "terminal symbols" which stand for themselves, like the letters in the filename patterns, and "non-terminal symbols", which name other rules. There are no examples of these above. It may also contain operators for combining patterns, such as the regular expression "|" and "*", or the Snobol "|", "any" and "arbno". The example below is a grammar for a fragment of English:

```
sentence      --> noun_phrase verb noun_phrase?
noun_phrase   --> proper_noun | determiner adjective* noun
proper_noun   --> "Mary" | "John" | "Fido"
noun          --> "apple" | "ball" | "cat" | "fish"
verb          --> "bites" | "eats" | "kicks" | "loves" | "sees"
adjective     --> "big" | "brown" | "small"
```

```
determiner    --> "the" | "a" | "each" | "every"
```

As in the patterns above, "|" indicates a choice between alternatives; "*" repeats an element indefinitely; "?" marks an optional component. The arrow "-->" means "is defined as". There are many possible notations: this is that used by Prolog Definite Clause Grammars.

**3.3 Spreadsheet grammars**

From these examples, it seems reasonable that a grammar for describing parts of spreadsheets would let us name and invoke grammatical rules, and would have operators for combining elements within these rules. It might not need terminal symbols, since, at least for general-purpose rules intended to apply to many spreadsheets, it's unlikely that we would want references to specific labels or formulae.

A big difference is dimensionality. Spreadsheets can go down a column and across worksheets as well as along a line; the grammar will need operators to state where one element lies relative to the next.

What about content? Suppose we have a spreadsheet laid out as follows (this one is from a project on modelling house prices). Each row consists of a label followed by a cell:

```
Household Income   cell
Interest Rate      cell
Population         cell
```

Using the same notation as before, we could say the rows follow the grammar:
```
row --> label cell
```

Now, suppose the spreadsheet had been in columns instead of rows:

```
Household Income   Interest Rate   Population
cell               cell            cell
```

We could now think in terms of columns, describing each by the rule:
```
column --> label DOWN cell
```
We have introduced a "DOWN" operator to indicate the need to go down rather than along.

Somewhat more complicated is our Income/Outgoings/Profit spreadsheet, where each column contains three cells headed by a label. We could describe its columns as:
```
column --> label DOWN cell DOWN cell DOWN cell
```
or, since programmers loath writing anything more than once if they can devise a way to automatically repeat it, we might introduce a "repeat N times" operator:
```
column --> label (DOWN cell)*3
```

This rule is actually a special case of a more general rule which applies to many spreadsheets and says that a column is a label with any number of cells beneath:

```
column --> label (DOWN cell)*
```
Once more, we use "*" without a right-hand argument to denote indefinite repetition.

Finally, how about the Property spreadsheet, the one that's split into blocks, with attributes splattered across the worksheet? There are several possible descriptions. One is to start with the blocks as structural units:

```
spreadsheet       --> (block ALONG(12))*3
block             --> label DOWN
                      (mortgage ALONG(2) other_costs) DOWN
                      rent DOWN profit
mortgage          --> cell
other_costs       --> cell
rent              --> cell
profit            --> cell
```
Here, we write an explicit ALONG operator, analogous to DOWN, rather than using implicit concatenation as we did in our first example. One advantage of this is that we can give the operator an argument, so that ALONG(12) means "go along 12 cells".

Another equally valid description would be to take the structural units to be attributes:

```
 spreadsheet         --> mortgage_parts AND other_costs_parts AND
                         rent_parts AND profit_parts
 mortgage_parts      --> DOWN              (cell ALONG(12))*3
 other_costs_parts   --> DOWN ALONG(3)     (cell ALONG(12))*3
 rent_parts          --> DOWN(2)           (cell ALONG(12))*3
 profit_parts        --> DOWN(3) ALONG(6)  (cell ALONG(12))*3
```

DOWN now has an argument too, for when we want to move down more than one row. We have also introduced an AND operator, which superimposes elements without moving.

What have we achieved in this section? We have adapted ideas from pattern-matching languages and Prolog DCG grammars to describe structure in spreadsheets. Some descriptions will apply to many spreadsheets; others, such as the Property ones, will be more specific. In the following section, we indicate how we implement this in Prolog.

## 4 LOGIC PROGRAMMING AND PROLOG

### 4.1 What is logic programming?

Logic programming languages are quite different from languages like C and Fortran in which one gives the computer a sequence of instructions - read data, assign to a variable, print variable - which it has to follow. In logic programming, by contrast, the programmer writes statements in logic that describe the properties that the solution must have.

By analogy, the C or Fortran programmer writing a program to build a house would code instructions telling the computer to pick up a brick, lay it on another brick, put a window next to it, and so on. The logic programmer, however, would write logical statements describing what it means for something to be a house - there are walls, which are parallel to one another and perpendicular to the ground, and composed of bricks in one of several patterns, and may contain a window, and other things. Compare how, later in this section, we use Prolog to find the labels in a spreadsheet. Our program is not a set of instructions, but a specification of what it means for a cell to be a label.

Logic programs consist of "facts" or logical statements. Some are unconditionally true; some depend on the truth of others. To run a logic program, one gives it a "goal": a logical statement which it must prove from the facts. We start with one of the simplest possible examples of Prolog, namely these two facts:

```
 mortal(X) :- man(X).
 man(socrates).
```

The first says that X is mortal if X is a man; the symbol ":-" means "if". In everyday language, all men are mortal. The second says Socrates is a man. So the first fact is conditionally true - it depends on whether its subject is a man - while the second is unconditionally true.

To run this program, we ask Prolog the goal:

```
 ?- mortal(socrates).
```

Prolog answers by finding the first fact, binding X to 'socrates', inferring that Socrates can be proven mortal if he can be proven a man, finding the second fact, thus proving him a man, and finally replying 'yes' to the goal.

More interesting are goals that don't merely check the truth of some statement, but find values for variables. If we ask

```
 ?- mortal(Y).
```
Prolog replies that Y=socrates.

### 4.2 Prolog

Prolog was the first widely used logic-programming language, and although others exist, it is still by far the most popular. There are some very good implementations around: we recommend SWI Prolog [SWI], which is efficient, free, and runs on Unix and Windows. This is what we are using in this project.

**4.3 Analysing spreadsheets in Prolog**

How can we apply logic programming to spreadsheets? In logic programming, we think in terms of predicates. "X is mortal" is a predicate we used above, as is "X is a man". Other predicates from everyday life include "X is inside Y", "X has colour C", "X costs P pence", "X is between Y and Z", "X is employed by B in department D at salary S", "the number I is less than the number J", "the set S contains element E". We see that many predicates express properties of a thing, or relations between things.

With spreadsheets, these things will include cells and cell ranges. We shall want to talk about properties of cells – "cell C contains formula F", "cell C is empty" – and relations between cells – "cell C depends on cell D", "cell C is in the same column as cell D", "cell C appears to be a copy of cell D", "cell C appears to be a label". We might also want predicates on bigger pieces of the spreadsheet – "column C appears to consist of cells which are all copies of one another", "the cells between (X1,Y1) and (X2,Y2) appear to be part of the same table". Note the relevance to structure discovery.

Some of the spreadsheet predicates will be derived directly from the spreadsheet. For example, the facts defining "cell C contains formula F" are just an enumeration of each cell's address and contents. As part of this project, we have written a Visual Basic program that dumps a spreadsheet in this form, so that it can be read into Prolog as a list of facts.

Other predicates will be defined in terms of these. The predicate "cell C depends on cell D" can be defined, roughly, like this:
  cell C depends on cell D if
    cell C contains formula F and
    F contains a subexpression SE and
    SE is a reference to D.

The predicate "cell C appears to be a label" can be defined as:
  cell C appears to be a label if
    cell C contains formula F and
    F is a string and
    there is no cell C' that depends on C.
where the final qualification allows us to exclude strings that are part of some calculation.

Given such definitions, we can ask Prolog to find values of variables that satisfy such predicates, thus finding (with the above examples) cells that depend on one another or are labels. We said that we might also want predicates on bigger pieces of the spreadsheet, such as "column C appears to consist of cells which are all copies of one another"; and we could equally well define and query these. We leave further discussion of Prolog to the Web version of our paper, which extends this section with a tutorial on its use for spreadsheet analysis. For a general introduction to Prolog, we recommend [Bratko, 1990] and [Sterling and Shapiro, 1994].

As far as structure discovery via grammars is concerned, we merely note that grammars are very closely related to predicates, and that logic programming has developed various techniques for implementing grammars by translating them into predicate definitions. These are well known amongst practitioners, and are introduced in the books just cited. We used them in the current project. Before we could do so however, we needed to invent data structures suitable for representing spreadsheets and their structure. That is the topic of the next section.

**5 ARROWS AND REVERSE ENGINEERING**

Before we could implement the previous section's structure-matching predicates, we needed a way to represent spreadsheets and the implicit structure we find. As Section 2.1 shows, MM gives us a way to do so. However, we still had to implement MM programs in Prolog, in a way that makes it easy to transform to and from spreadsheets, search for structure, and retain information about spreadsheet layouts. This section, which requires slightly more maths than the rest of the paper, describes how we did this. We believe it is an elegant and concise way of formulating reverse engineering for spreadsheets.

Let us begin by imagining one worksheet of a spreadsheet in value view. This is, in effect, an array whose indices are cell addresses and whose elements are values, i.e. numbers, strings, and other data. Mathematically, we can consider it a function from cell addresses to values.

Value view is boring, so let's switch to formula view. Now, each cell maps to a formula or expression, and the worksheet becomes a function from cell addresses to expressions

MM views the world as a collection of attributes, each attribute being an array. These too can be considered functions, from the attribute's indices to MM expressions.

An MM program, then, is modelled as a collection of functions, one for each attribute. Mathematically, we regard it as an indexed function, the indices being the attributes.

Compiling an MM program translates it into a spreadsheet; i.e., translates this indexed function into a single function from cell addresses to Excel expressions.

Now, before we can compile an MM program, we need to know how each attribute maps onto the spreadsheet. In our Income/Outgoings/Profit spreadsheet, each attribute mapped onto a column; in the Property spreadsheet, the attributes (Rent, Profit and so on) mapped onto cells staggered across the worksheet. In general, the compiler will assume a default mapping for each attribute (it allocates attributes to columns headed by their name), but this can be overridden by the user. In any case, MM clearly needs to know what the mapping from attribute to spreadsheet will be.

We now see that compiling an MM program becomes a matter of taking the attribute-to-expression functions, pre-composing with the attribute-to-spreadsheet mappings, and forming the union of the results. To generate a spreadsheet output file, we iterate over the domain of this union arrow, pumping out each element and its image - i.e. the cell address and its contents.

Decompilation goes in the opposite direction, from the spreadsheet-to-expression function to a set of attribute-to-expression functions. The trial and error of spreadsheet algebra entails finding a decomposition optimal for intelligibility; structure discovery infers appropriate domains for the attribute functions. In general, there is no unique decomposition, hence the need for trial, error, and human assistance.

Our latest version of MM takes this model literally, and represents the functions as Prolog data structures. We note for Prolog programmers that we found this good for manipulating functions in a representation-independent way, without needing to worry about whether they are stored as facts (clauses) or as association lists, trees, and so on.

Finally, many branches of mathematics like to consider functions as "arrows". Amongst other things, this gives us, via "commutative diagrams", a handy graphical way to depict them, and enables some proofs to be done largely by drawing. We found this very useful, and think likewise of our functions as arrows.

**6 IMPLEMENTATION**

As mentioned in Section 4, the current MM is coded in SWI Prolog. Advantages over Kawa, in which we coded the previous version [Paine, 2004], include: easier to generate .exe files for; probably less memory-heavy; the syntax is more readable, allowing one to define new operators, so making it easy to prototype notations for patterns without writing a parser; the TLI's query interface also helps rapid prototyping. And logic programming is wonderful. SWI Prolog may not be as portable as the Java-based Kawa, but it does run on all the machines we've needed so far.

**6.1 The user interface**

Ideally, our system would just ask the user for a spreadsheet, and would then completely automatically match it against a library of structure-discovery rules, producing that MM program which best reveals the structure implicit in the spreadsheet. However, such complete automation is unlikely to be feasible. The user will often be able to contribute additional information about structure, or may want to guide the system toward a particular form for the program. This is why we developed the trial-and-error computer-algebra style interface of [Paine, 2004].

The present system will implement the same interface, together with a means of entering structure-discovery patterns which can be matched against a spreadsheet to suggest how cells may be grouped into attributes, and what names these attributes might have. The suggestions will be data structures which can be passed as arguments to the spreadsheet algebra functions. So, for example, if we are working on a particular spreadsheet, we might match it against our patterns and be told that the non-empty cells in column C all seem to be part of one attribute, named "Profit". We could then pass that information to a function that merges these cells into the designated attribute, renames all references, and displays the result.

We are starting with a text-only interface, based on Prolog's top-level interpreter. This is the part of a Prolog implementation to which the user gives Prolog goals to prove, getting back their truth or falsity, plus the values Prolog finds for variables. We are augmenting this with our Grips pre-processor [Paine, 2003], which allows goals to be written as function calls rather than predicate invocations. This makes certain kinds of goal more compact and easier to write.

Once we have the text-only interface working, we shall prototype a graphical user interface. We shall do this as a Web application where MM will run on a Web server, the user operating it via a browser. This is less interactive than a windowing interface, but easier to prototype, since the browser takes over the grunt work of rendering. We have taken advantage of this before in our Spreadsheet Autopublisher [Paine; Paine and Ramsden, 2002], a Web-based application which automatically converted simple spreadsheets or MM programs submitted to it into other applications runnable over the Web.

**6.2 Specificity versus generality**

We have to decide how much the user is allowed to do. Can they invoke any Prolog predicate on the spreadsheet, even defining their own if they know how? Or do we restrict them in some way?

We shall certainly provide access directly to Prolog, if only because we shall need it ourselves when debugging. However, although Prolog is closer to logic than many other languages, it is not a complete implementation, and Prolog programmers need to learn a number of non-obvious tricks before they can write predicates that are guaranteed not to loop indefinitely, generate alternative answers that are never needed, or handle negation incorrectly. To avoid users having to become skilled Prolog programmers, we will therefore provide a library of elementary predicates together with operations for combining these, and perhaps some limited facilities for defining new ones. The combining operations will be built on top of Prolog's built-in logical operators, in much the same way that the AND, DOWN and ALONG operators in our spreadsheet grammars could be. This will enable us to hide some of the messy details of Prolog from the user, and will also give us the freedom to reorder patterns to make them more efficient.

**6.3 Efficiency**

This brings us to efficiency. In general, our approach is "make the program correct first, then elegant, then efficient". There are too many languages and systems around – perhaps Java is one – designed with no thought for fundamentals. And computers are still getting faster. I can now carry more computing power in my rucksack – complete with SWI Prolog compiler – than Oxford University had to share between all my students when I first started teaching Prolog.

Having said this, we do need to decide what to do if our pure-Prolog approach is infeasibly slow. One point is that many Prologs contain features to speed up the search for appropriate facts – "clause indexing", as Prolog programmers know it. For example, Dennis Merrit of Amzi! [Amzi!] has told us that Amzi! Prolog implements indexed dynamic database predicates using B-trees, and that this gives them very good response times with the large WordNet libraries (WordNet is an online dictionary which models some features of human linguistic memory), which have 170,000 clauses in the larger predicates.

If such features turn out to be inadequate, we may be able to implement our own. All decent Prologs have a "foreign language" interface, allowing them to call code written in C and other languages. By using this, we could define our elementary predicates in C (say), optimising them to the hilt but still making them look to the user like Prolog, so that the non-logic-programming aspects are safely hidden. We have done this before when teaching Prolog. One of our students wanted to write a limerick generator, whose output would rhyme, scan, and even be grammatically correct. We had a machine-readable dictionary of about 70,000 words, whose files gave the pronunciation, scansion, and parts of speech for each word. We converted this into an indexed file, wrote some indexed-file-search routines in Fortran, and then interfaced them to our Prolog system (which in those days was part of Sussex Poplog) using Pop-11. The result was that the student could ask Prolog goals such as "find all nouns which rhyme with 'chimney'", or "find all verbs which are 8 syllables long". The goals behaved as if defined in Prolog, but were actually doing a fast search implemented by the operating system's indexed sequential files.

### 6.4 Progress so far

The most important part of our work was getting the representations described in Section 5 (Arrows and Reverse Engineering) right, as this gives us a proper mathematical foundation for everything else. That is now done, and we have implemented the representation in Prolog and tested it with a variety of structure-discovery patterns and spreadsheet algebra functions, using a Visual Basic program to dump spreadsheets as Prolog facts.

Given that we have the Prolog top-level interpreter and our Grips pre-processor (Section 6.1), most of the text-only interface exists, although we are still deciding on the most appropriate elementary predicates to give the user. We have also started work on the Web interface, which we are implementing as PHP [PHP] scripts that call SWI-Prolog.

We have not yet hit any efficiency problems – in our tests, the largest spreadsheet has had about 100 occupied cells out of 200. However, as mentioned above, we do have some ideas on how to attack such problems should they arise.

### 7 FUTURE WORK

We need to experiment with alternative ways to write patterns. Once the notation is good enough to be fixed, we would like to see researchers collaborate in building up a shared library of structure patterns.

### 7.1 Intelligent structure monitoring and flexible best practice

Our goal is to run MM as an "intelligent structure monitor" alongside Excel. This could revolutionise spreadsheet "best practice". We have said already that different spreadsheet tasks require different layouts. A layout convenient to a user entering data and reading off results may be much less convenient to the maintenance programmer. The problem is that all guidelines for best-practice demand a single layout. But no layout can be simultaneously optimum for tasks as different as data entry and maintenance. The answer is to move from a view of spreadsheets as single unchanging entities to a relativistic approach where they can freely adapt to the user's needs.

MM will provide this.

## 8 ACKNOWLEDGEMENTS

We would like to thank one of the anonymous referees for comments, and the Excelsior café in Cowley for coffee.